\documentclass[11pt]{article}
\usepackage[a4paper,margin=1in]{geometry}

\usepackage{tikz}
\usetikzlibrary{arrows.meta} 
\usetikzlibrary{shadows}     
\usetikzlibrary{shapes}      
\usetikzlibrary{3d, calc, backgrounds}
\usetikzlibrary{decorations.markings}
\usepackage{pgfplots}
\pgfplotsset{compat=1.18}
\usepackage{tikz-3dplot}
\definecolor{flowblue}{RGB}{50, 100, 200}
\definecolor{vortred}{RGB}{220, 50, 50}
\definecolor{metricorange}{RGB}{230, 140, 30}

\definecolor{deepblue}{RGB}{0, 60, 140}
\definecolor{cyanflow}{RGB}{0, 160, 200}
\definecolor{fireorange}{RGB}{255, 100, 0}
\definecolor{crimson}{RGB}{220, 20, 60}
\definecolor{softgray}{RGB}{240, 240, 245}
\definecolor{gridline}{RGB}{180, 180, 180}
\usepackage{amsmath,amsthm,mathtools}

\usepackage{newtxtext,newtxmath}

\usepackage{microtype}
\usepackage{bm}
\usepackage{graphicx}

\usepackage{hyperref}
\hypersetup{
    colorlinks=true,      
    linkcolor=blue,       
    citecolor=blue,       
    urlcolor=blue,        
    filecolor=blue        
}

\usepackage{cleveref}

\theoremstyle{definition}

\theoremstyle{remark}

\title{\textbf{Lagrangian Phase-Lag and Geometric Precedence in Turbulent Vortex Stretching}}
\author{
\textbf{Khalid M. Saqr}\\
\small Mechanical Engineering Deparatment, College of Engineering and Technology \\
\small Arab Academy for Science, Technology and Maritime Transport\\ \small Main Campus, Alexandria (1029)--EGYPT\\
\small ORCID: \href{https://orcid.org/0000-0002-3058-2705}{0000-0002-3058-2705}
}

\date{}

\begin{document}
\maketitle
\begin{abstract}
\noindent This study investigates the causal timeline of vortex stretching in high-Reynolds-number turbulence ($Re_\lambda \approx 433$) using Lagrangian tracking in $1024^3$ direct numerical simulations. While classical theories often assume an instantaneous alignment between strain and vorticity, the present analysis identifies a systematic Lagrangian phase lag governing the onset of intense dissipation. By conditionally averaging the dynamics of fluid parcels, a distinct phase-space hysteresis is revealed. Trajectories are captured by the saddle-point topology of the pressure field ($\lambda_{min}^p < 0$) \textit{prior} to experiencing peak enstrophy amplification. This temporal ordering ($\tau > 0$) demonstrates that the pressure topology acts as a deterministic geometric precursor, organizing the flow structure before the bursting event occurs. The robustness of this mechanism is verified in magnetohydrodynamic (MHD) turbulence, where the Lorentz force is found to suppress the hysteresis loop, forcing a transition from causal precedence to simultaneous locking.
\end{abstract}


\section*{Introduction}

The formation of small scales in three-dimensional turbulence is driven by the stretching of vortex lines, a mechanism that remains central to the study of the Euler and Navier--Stokes equations \cite{majda}. A longstanding challenge in turbulence modeling is identifying exactly \textit{where} and \textit{why} this stretching becomes singular. Classically, this problem is approached in the Eulerian frame, where the vorticity transport equation, $D\boldsymbol{\omega}/Dt = (\boldsymbol{\omega} \cdot \nabla)\mathbf{u}$, relates stretching to the alignment between vorticity and the eigenvectors of the rate-of-strain tensor, $S$ \cite{Ohkitani1993, Tsinober2001}. While this view successfully describes instantaneous dynamics, it does not easily capture the ``history'' of deformation. A fluid parcel entering a region of high strain may not necessarily experience significant stretching if it rotates out of alignment too quickly.

To capture this history, we turn to geometric hydrodynamics. Following the foundational insights of Arnold \cite{arnold}, fluid motion can be viewed as geodesic flow on the group of volume-preserving diffeomorphisms. In this Lagrangian framework, instability is not merely an instantaneous alignment, but a cumulative divergence of trajectories encoded in the \textit{pullback metric} (the right Cauchy--Green tensor), $C_t$ \cite{haller2015}. In this work, the author aims to bridge the gap between the Eulerian pressure field and Lagrangian metric deformation. The widely studied ``vortex stretching'' is found, as shown in later sections, to be geometrically equivalent to the evolution of the pullback metric driven by the \emph{Pressure Hessian}, $\nabla \nabla p$. Specifically, the Pressure Hessian is found to play the role of an effective curvature operator governing the stability of fluid trajectories.

This geometric perspective builds upon a rich history of examining the nonlocal role of pressure. The seminal early work by Ohkitani \cite{Ohkitani1993, Ohkitani1995} and Nomura \& Post \cite{Nomura1998} identified that in regions of maximum enstrophy, the vorticity vector tends to align with the eigenvectors of the Pressure Hessian, $\nabla\nabla p$. This phenomenon suggests a strong ``geometric locking'' between the non-local pressure field and local stretching dynamics.The precise dynamical role of this locking has been recently debated. Buaria et al. \cite{Buaria2022} argued that the Pressure Hessian primarily acts to deplete intense strain, redistributing fluctuations toward the mean field. Conversely, Yang et al. \cite{Yang2024} demonstrated that the anisotropic component of the Pressure Hessian is structurally essential for enabling strong vortex stretching, distinguishing it from isotropic models that inhibit growth. 
However, these studies primarily focus on Eulerian correlations \cite{Yu2021} or instantaneous Lagrangian alignments \cite{Tom2021}. The fundamental question of \emph{temporal ordering} remains unanswered: does the pressure topology act as a deterministic architect that prepares the flow for instability, or is it merely a simultaneous byproduct of the stretching process itself?

To resolve this ambiguity, the proposed mechanism is tested against high-resolution datasets from the Johns Hopkins Turbulent Databases (JHTDB) \cite{Li2008, Perlman2007}. By analyzing forced isotropic turbulence ($Re_\lambda \approx 433$) and comparing it with magneto-hydrodynamic (MHD) flows, we examine the universality of this geometric constraint. Specifically, a Lagrangian time-series analysis is employed to determine the phase-lag between the formation of negative pressure curvature ($\lambda_{min}^p < 0$) and the subsequent peak in enstrophy production. The transition from a purely Eulerian view of strain alignment to a Lagrangian view of pressure-driven metric instability provides a coherent geometric criterion for the onset of turbulence.

The paper is organized as follows. Section 2 reviews and discusses the geometric transport of vorticity and derives the main result, identifying the Pressure Hessian as the driver of metric anisotropy in a Jacobi-style stability equation. Section 3 provides a comparative analysis of the JHTDB isotropic and MHD datasets, focusing on the time-lagged cross-correlation and the alignment of the vorticity vector within the pressure scaffold. Finally, Section 4 discusses the implications for turbulence theory and singularity criteria.

\section*{Theoretical Framework and Mathematical Derivation}

Let $(M,g)$ be a compact, oriented, three-dimensional Riemannian manifold representing the fluid domain. The velocity field $\mathbf{u}$ generates a flow map $\varphi_t: M \to M$, defined by the differential equation $\frac{d}{dt} \varphi_t(X) = \mathbf{u}(\varphi_t(X))$ with initial condition $\varphi_0(X) = X$. The Lagrangian deformation of fluid elements is encoded in the deformation gradient $F = D\varphi_t$. The evolution of the manifold's geometry is tracked via the \emph{right Cauchy--Green tensor}, $C_t(X)$, defined as the pullback of the ambient metric $g$ by the flow:
\begin{equation}
C_t(X) := \varphi_t^* g|_{\varphi_t(X)}, \quad (C_t)_{ij} = \frac{\partial \varphi^k}{\partial X^i} \frac{\partial \varphi^l}{\partial X^j} g_{kl}.
\end{equation}
Physically, $C_t$ acts as a time-dependent metric on the initial configuration, where the squared length of a material line element $dX$ at time $t$ is given by $C_t(dX, dX)$.

\subsection*{Vorticity as a Lie-Transported Two-Form}
To utilize the intrinsic geometry of the flow, the velocity $\mathbf{u}$ is identified with its metric dual one-form $u^\flat = g(u, \cdot)$. The vorticity is defined as the exact two-form $\boldsymbol{\omega} = d u^\flat$. In inviscid Euler flows, the vorticity two-form is Lie-transported by the flow, $\mathcal{L}_u \boldsymbol{\omega} = 0$ \cite{arnold, marsden}. For the associated vector field $\omega^\sharp$, this yields the Cauchy invariant:
\begin{equation}
\omega(x,t) = (\varphi_{t*} \omega_0)(x) = D\varphi_t(X) \cdot \omega_0(X).
\end{equation}
This geometric identity implies that the vorticity vector behaves exactly like a tangent vector $dX$ under the flow. Consequently, the evolution of enstrophy is a direct manifestation of the metric distortion of the fluid manifold.

\subsection*{Vorticity Stretching as Metric Distortion}
To quantify the cumulative history of deformation, we utilize the geometric identity relating enstrophy to the pullback metric. The squared norm of the vorticity vector along a trajectory is given exactly by the contraction of the initial vorticity with the time-dependent metric tensor $C_t$:
\begin{equation}
|\omega(t)|_g^2 = g(\varphi_{t*} \omega_0, \varphi_{t*} \omega_0) = (\varphi_t^* g)(\omega_0, \omega_0) = C_t(\omega_0, \omega_0).
\label{eq:metric_norm}
\end{equation}
Differentiating \eqref{eq:metric_norm} yields the familiar Eulerian stretching term, $2 \boldsymbol{\omega} \cdot S \boldsymbol{\omega} = \dot{C}_t(\omega_0, \omega_0)$. While this relationship is kinematic, it provides a crucial Lagrangian advantage: it allows us to decompose the instantaneous stretching rate into a cumulative geometric path history.

\subsection*{The Jacobi-style Deviation Equation and the Pressure Hessian}
The geometric instability of the flow is localized by analyzing the relative acceleration of vorticity through the second Lagrangian time derivative of the deformation gradient $F$. The velocity gradient $\nabla \mathbf{u}$ evolves according to the matrix Riccati equation:
\begin{equation}
\frac{D (\nabla \mathbf{u})}{Dt} = - \nabla \nabla p - (\nabla \mathbf{u})^2.
\end{equation}
Taking the Lagrangian time derivative of the evolution equation $\frac{DF}{Dt} = (\nabla \mathbf{u})F$ yields:
\begin{equation}
\frac{D^2 F}{Dt^2} = \left( \frac{D (\nabla \mathbf{u})}{Dt} + (\nabla \mathbf{u})^2 \right) F.
\label{eq:F_accel}
\end{equation}
Substituting the Riccati equation into \eqref{eq:F_accel} results in the exact cancellation of the nonlinear advection terms $(\nabla \mathbf{u})^2$. This leaves a linear system defining the Lagrangian acceleration:
\begin{equation}
\frac{D^2 F}{Dt^2} + (\nabla \nabla p) F = 0.
\end{equation}
Since $\boldsymbol{\omega}(t) = F(t) \cdot \boldsymbol{\omega}_0$, the vorticity evolution is governed by the Jacobi-style deviation equation:
\begin{equation}
\frac{D^2 \boldsymbol{\omega}}{Dt^2} + \mathcal{K}(t) \boldsymbol{\omega} = 0, \quad \text{where } \mathcal{K}(t) = \nabla \nabla p.
\label{eq:jacobi_vort}
\end{equation}
In this framework, the Pressure Hessian acts as the effective curvature operator. Enstrophy blow-up is dynamically linked to the "effective negative curvature" ($\lambda_{min}^p < 0$) provided by the saddle-point topology of the pressure field. 

Equation~\eqref{eq:jacobi_vort} is derived exactly for inviscid Euler flows, where the vorticity two-form is Lie-transported. In the Navier--Stokes equations, the presence of the viscous term $\nu \Delta \boldsymbol{\omega}$ breaks exact Lie transport and introduces an additional contribution to the second material derivative of $\boldsymbol{\omega}$. Formally, this results in a modified deviation equation of the form
\begin{equation}
\frac{D^2 \boldsymbol{\omega}}{Dt^2} + \nabla\nabla p\,\boldsymbol{\omega}
= \nu\,\mathcal{R}(\boldsymbol{\omega},\nabla\boldsymbol{\omega},\nabla^2\boldsymbol{\omega}),
\label{eq:viscous_jacobi}
\end{equation}
where $\mathcal{R}$ denotes residual terms involving higher-order spatial derivatives. It is important to clarify the physical scope of this equation. We do not claim that the inviscid geometric instability leads to a finite-time singularity in the Navier--Stokes equations; viscous diffusion ($\nu \mathcal{R}$) will inevitably regularize the flow at the Kolmogorov scale. However, our focus is on the \emph{onset} of intense stretching within the inertial range. In this regime, the dynamics are dominated by inertia, and the Pressure Hessian acts as the primary driver of geodesic deviation before viscous terms become order-one. The DNS analysis therefore tests whether this inviscid geometric scaffold correctly predicts the \textit{timing} of enstrophy amplification, even if viscosity ultimately limits its magnitude.

The dominance of the geometric instability over viscous damping is determined by the ratio of the curvature timescale $\tau_{geo} \sim |\lambda_{min}^p|^{-1/2}$ to the viscous diffusion timescale $\tau_{\nu} \sim \delta^2/\nu$ across the stretching event. In the inertial range, where vortex stretching is initiated, the dynamics are dominated by inertia such that $\tau_{geo} \ll \tau_{\nu}$. Consequently, while viscosity eventually halts the singularity, the \textit{onset} and \textit{orientation} of the stretching are governed by the inviscid geometric scaffold. The DNS analysis therefore tests whether this inviscid geometric trigger remains the predictive driver of instability even in the presence of dissipative regularization.

To investigate this mechanism in high-Reynolds number turbulence, we utilize the Johns Hopkins Turbulent Databases (JHTDB). The requirement for $C^2$ continuity in \eqref{eq:jacobi_vort} necessitates the use of 8-point Hermite spline interpolation (\texttt{m2q8}) to extract the Pressure Hessian along Lagrangian trajectories. This ensures that the extracted curvature $\mathcal{K}(t)$ is independent of grid-scale noise, permitting a deterministic assessment of temporal causality across hydrodynamic and magneto-hydrodynamic (MHD) regimes.

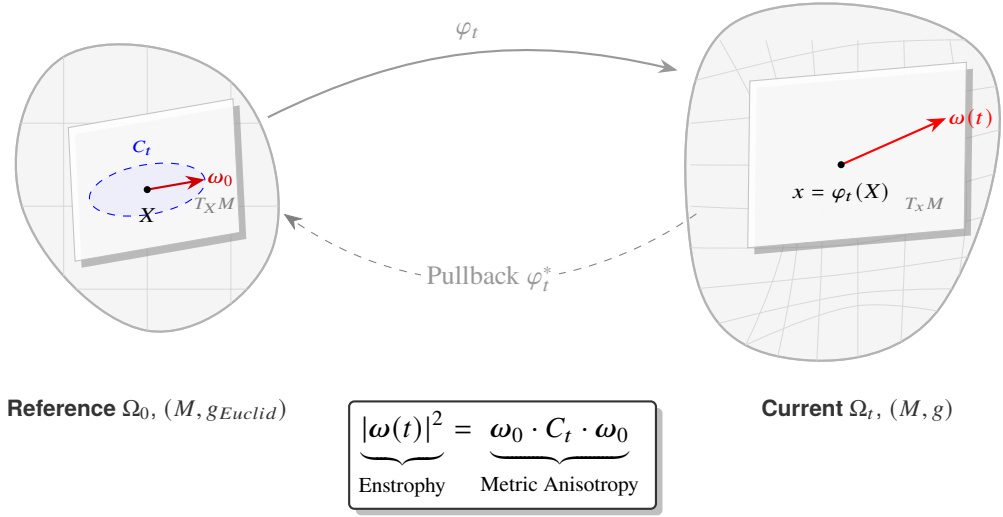
\begin{figure}[t]
\centering
\begin{tikzpicture}[
    scale=1.1,
    >={Stealth[length=2.5mm]},
    manifold/.style={thick, fill=gray!8, draw=gray!40, line join=round},
    gridlines/.style={draw=gray!30, ultra thin},
    tangentplane/.style={fill=white, draw=gray!50, opacity=0.9, thin, drop shadow},
    vector/.style={->, thick, line cap=round},
    annotation/.style={font=\footnotesize\sffamily, text=black!80},
    mathlabel/.style={font=\footnotesize, text=black}
]

    
    \path[manifold] plot[smooth cycle, tension=0.7] coordinates {(-2.5,0) (-2,1.8) (0,1.5) (0.5,-0.5) (-1,-1.5)};
    \draw[thick, gray!60] plot[smooth cycle, tension=0.7] coordinates {(-2.5,0) (-2,1.8) (0,1.5) (0.5,-0.5) (-1,-1.5)};
    
    \begin{scope}
        \clip plot[smooth cycle, tension=0.7] coordinates {(-2.5,0) (-2,1.8) (0,1.5) (0.5,-0.5) (-1,-1.5)};
        \draw[gridlines] (-3,-2) grid (1,2);
    \end{scope}

    \coordinate (X) at (-1, 0.2);
    \begin{scope}[shift={(X)}, rotate=10]
        \draw[tangentplane] (-1.0,-0.7) -- (1.0,-0.7) -- (1.2,0.9) -- (-0.8,0.9) -- cycle;
        \node[font=\tiny, gray, anchor=south east] at (1.1, -0.6) {$T_X M$};
        
        \fill[blue!10, opacity=0.5] (0,0) ellipse (0.7cm and 0.3cm);
        \draw[blue, dashed, thin] (0,0) ellipse (0.7cm and 0.3cm);
        \node[blue, font=\tiny, anchor=south] at (0, 0.3) {$C_t$};
        
        \draw[vector, red!80!black] (0,0) -- (0.7,0) node[right, inner sep=1pt, font=\scriptsize] {$\bm{\omega}_0$};
        \fill[black] (0,0) circle (1.2pt);
    \end{scope}
    
    \node[below=3pt, font=\scriptsize] at (X) {$X$};
    \node[annotation, anchor=north] at (-1,-2.2) {\textbf{Reference} $\Omega_0$, $(M, g_{Euclid})$};

    
    \begin{scope}[shift={(7.5,0)}] 
        
        \path[manifold] plot[smooth cycle, tension=0.7] coordinates {(-2, -0.5) (-1.5, 2.2) (1.5, 1.8) (1, -1.2) (-1, -1.8)};
        \draw[thick, gray!60] plot[smooth cycle, tension=0.7] coordinates {(-2, -0.5) (-1.5, 2.2) (1.5, 1.8) (1, -1.2) (-1, -1.8)};

        \begin{scope}
            \clip plot[smooth cycle, tension=0.7] coordinates {(-2, -0.5) (-1.5, 2.2) (1.5, 1.8) (1, -1.2) (-1, -1.8)};
            \foreach \x in {-1.5,-1,...,1.5}
                \draw[gridlines] plot[smooth, tension=0.6] coordinates {(\x, -2) (\x + 0.3*\x*\x, 0) (\x + 0.1*\x, 2)};
            \foreach \y in {-2,-1.5,...,2.5}
                \draw[gridlines] plot[smooth, tension=0.6] coordinates {(-2, \y) (0, \y + 0.2*\y*\y) (2, \y + 0.5)};
        \end{scope}

        \coordinate (x_curr) at (-0.2, 0.5);
        \begin{scope}[shift={(x_curr)}, rotate=5]
            \draw[tangentplane] (-1.2,-0.9) -- (1.4,-0.9) -- (1.6,1.1) -- (-1.0,1.1) -- cycle;
             \node[font=\tiny, gray, anchor=south east] at (1.3, -0.8) {$T_x M$};
             
            \draw[vector, red] (0,0) -- (1.3, 0.45) node[right, inner sep=1pt, font=\scriptsize] {$\bm{\omega}(t)$};
            \fill[black] (0,0) circle (1.2pt);
        \end{scope}
        
        \node[below=3pt, font=\scriptsize] at (x_curr) {$x = \varphi_t(X)$};
        \node[annotation, anchor=north] at (0,-2.2) {\textbf{Current} $\Omega_t$, $(M, g)$};
    \end{scope}

    
    \draw[->, thick, gray!80, shorten >= 5pt, shorten <= 5pt] 
        (0.3, 1.0) to[bend left=20] 
        node[midway, above=2pt, font=\small] {$\varphi_t$} 
        (5.5, 1.6);

    \draw[dashed, gray, ->, shorten >= 5pt, shorten <= 5pt] 
        (5.7, 0.0) to[bend left=35] 
        node[midway, fill=white, inner sep=2pt, font=\small, text=gray!80] {Pullback $\varphi_t^*$} 
        (0.5, 0.0);

    \node[draw=black!80, thick, rounded corners=2pt, fill=white, drop shadow, align=center] at (3.25, -3.0) {
        $\displaystyle \underbrace{|\bm{\omega}(t)|^2}_{\text{Enstrophy}} = \underbrace{\bm{\omega}_0 \cdot C_t \cdot \bm{\omega}_0}_{\text{Metric Anisotropy}}$
    };

\end{tikzpicture}
\caption{\textbf{Geometric Interpretation of the Lagrangian Stability Analysis.} The flow map $\varphi_t$ acts as a time-dependent coordinate transformation. Rather than tracking the vorticity vector $\boldsymbol{\omega}$ in a fixed frame, we track the deformation of the underlying metric tensor $C_t$ (blue ellipse). The "Geometric Locking" hypothesis tested in this paper posits that the Pressure Hessian $\nabla \nabla p$ determines the growth rate of the principal axis of $C_t$ (the unstable manifold) \textit{before} the vorticity vector fully aligns with it.}
\label{fig:metric_pullback}
\end{figure}

\subsection*{Illustrative Example: 3D Linear Stagnation Flow}

To explicitly visualize the geometric mechanism derived in the preceding sections, 3D linear stagnation flow is analyzed. While elementary, this exact solution serves as a canonical local model for the stretching of a vortex filament by a larger-scale strain field, permitting the derivation of the metric tensor $C_t$ and its growth rate in closed form.

The velocity field is considered in Cartesian coordinates $(x,y,z)$:
\begin{equation}
\mathbf{u}(x,y,z) = (-\alpha x, -\beta y, (\alpha + \beta)z), \quad \alpha, \beta > 0.
\end{equation}
The solenoidal condition $\nabla \cdot \mathbf{u} = 0$ is satisfied. For simplicity, parameters are set to $\alpha = \beta = 1$, representing axisymmetric compression in the $xy$-plane and stretching along the $z$-axis:
\[
\mathbf{u} = (-x, -y, 2z).
\]
The rate-of-strain tensor $S$ is constant and diagonal, given by $S = \text{diag}(-1, -1, 2)$. The flow map $\varphi_t(X_0, Y_0, Z_0)$ is obtained by integrating the velocity field:
\begin{align*}
x(t) &= X_0 e^{-t}, \\
y(t) &= Y_0 e^{-t}, \\
z(t) &= Z_0 e^{2t}.
\end{align*}
The deformation gradient $F = D\varphi_t$ is identified as:
\begin{equation}
F(t) = \begin{pmatrix}
e^{-t} & 0 & 0 \\
0 & e^{-t} & 0 \\
0 & 0 & e^{2t}
\end{pmatrix}.
\end{equation}

The right Cauchy--Green tensor $C_t = F^\top F$ is derived as:
\begin{equation}
C_t = \begin{pmatrix}
e^{-2t} & 0 & 0 \\
0 & e^{-2t} & 0 \\
0 & 0 & e^{4t}
\end{pmatrix}.
\end{equation}
This tensor reveals the extreme anisotropy of the metric, where the $z$-direction represents the unstable manifold characterized by exponential growth in metric costs. 

To evaluate the vorticity evolution, an initial vorticity perturbation aligned with the stretching direction is considered: $\boldsymbol{\omega}_0 = (0, 0, \Omega_0)$. 

\noindent \textbf{Dynamical Calculation (Eulerian):}
The stretching term in the vorticity transport equation is evaluated as $(\boldsymbol{\omega} \cdot \nabla)\mathbf{u} = 2\Omega \hat{k}$, leading to exponential growth $\Omega(t) = \Omega_0 e^{2t}$ and enstrophy density $|\boldsymbol{\omega}(t)|^2 = \Omega_0^2 e^{4t}$.

\begin{figure}[t!]
\centering
\tdplotsetmaincoords{70}{110} 
\begin{tikzpicture}[tdplot_main_coords, scale=2.0]
    \tikzset{
        >=latex,
        font=\small,
        axis/.style={->, black!60, thick},
        grid/.style={gridline, very thin, dashed},
        sphere_init/.style={fill=blue!5, opacity=0.3, draw=blue!20},
        ellipsoid_final/.style={left color=fireorange!80, right color=fireorange!40, middle color=fireorange!20, opacity=0.85, draw=crimson, thin},
        vort_arrow/.style={crimson, ultra thick, -{Latex[length=3mm, width=2mm]}},
        label_pin/.style={pin distance=0.8cm, pin edge={black!40, thin, -}}
    }

    \begin{scope}[xshift=-2.0cm] 
        \node[anchor=south, font=\bfseries] at (0,0,3.2) {(a) Flow Field $\bm{u}(\bm{x})$};
        \begin{scope}[canvas is xy plane at z=0]
            \draw[grid] (-1.5,-1.5) grid (1.5,1.5);
            \foreach \x in {-1.5, 1.5} \draw[->, deepblue, thick] (\x,0) -- (\x*0.4,0);
            \foreach \y in {-1.5, 1.5} \draw[->, deepblue, thick] (0,\y) -- (0,\y*0.4);
        \end{scope}
        \draw[axis] (0,0,0) -- (0,0,2.8) node[left] {$z$};
        \foreach \ang in {0, 90, 180, 270} {
            \tdplotsetrotatedcoords{\ang}{0}{0}
            \begin{scope}[tdplot_rotated_coords]
                \draw[cyanflow, thick, ->] (1.6, 0, 0.2) to[out=180, in=270] (0.6, 0, 0.9);
                \draw[fireorange, thick, ->] (0.6, 0, 0.9) to[out=90, in=270] (0.3, 0, 2.5);
            \end{scope}
        }
        \node[align=right, deepblue] at (-1.6, -2.1, 0) {Inflow\\(Compression)};
        \node[align=left, crimson] at (2.0, 0, 3) {Outflow\\(Stretching)};
    \end{scope}

    \begin{scope}[xshift=2.0cm] 
        \node[anchor=south, font=\bfseries] at (0,0,3.2) {(b) Material Deformation};
        \draw[axis] (0,0,0) -- (1.6,0,0) node[anchor=north east]{$x$};
        \draw[axis] (0,0,0) -- (0,1.6,0) node[anchor=north west]{$y$};
        \draw[axis] (0,0,0) -- (0,0,2.8) node[left]{$z$};
        \shade[ball color=gray!10, opacity=0.2] (0,0,0) circle (0.85cm);
        \draw[gray!30, dashed] (0,0,0) circle (0.85cm);
        \draw[gray!30, very thin] (0.85,0,0) arc (0:180:0.85 and 0.25);
        \def\S{2.3} \def\R{0.5} 
        \fill[ellipsoid_final] plot[domain=0:360, samples=40] ({\R*cos(\x)}, {\R*sin(\x)}, \S) 
            -- plot[domain=360:0, samples=40] ({\R*cos(\x)}, {\R*sin(\x)}, 0) -- cycle;
        \shade[left color=fireorange!90, right color=fireorange!30, middle color=fireorange!60, opacity=0.7] 
             (-\R, 0, 0) to[out=90, in=270] (-\R*0.4, 0, \S) 
             -- (\R*0.4, 0, \S) to[out=270, in=90] (\R, 0, 0) -- cycle;
        \fill[crimson!80] (0,0,\S) ellipse ({\R*0.4} and {\R*0.15});
        \draw[crimson, dashed, thin] (0,0,0) ellipse ({\R} and {\R*0.3});
        \draw[vort_arrow] (0,0,0) -- (0,0,\S+0.6) node[above] {$\bm{\omega}(t)$};
        \draw[<->, black!80] (1.0, 0, 0) -- (1.0, 0, \S);
        \node[right, black!80, font=\footnotesize] at (1.0, 0, \S/2) {Stretch $\lambda_1 > 0$};
        \draw[->, black!80] (0, -1.6, 0) -- (0, -0.9, 0);
        \node[below, black!80, font=\footnotesize, align=center] at (0, -1.6, 0) {Compress $\lambda_2 < 0$};
        \node[coordinate] (tip) at (0,-0.1,\S) {};
        \node[pin={[pin edge={black, thin}]100:{\color{crimson}\textbf{Vorticity aligns with Extension}}}] 
            at (tip) {};
    \end{scope}
\end{tikzpicture}
\caption{\textbf{Geometric structure of 3D linear stagnation flow.} \textbf{(a)} The Eulerian flow field $\bm{u}(\bm{x})$. Streamlines indicate compression in the horizontal $xy$-plane (blue inflow) and extension along the vertical $z$-axis (orange outflow). \textbf{(b)} Lagrangian material deformation. An initial unit sphere (gray ghost) deforms into a prolate ellipsoid (solid orange) due to volume-conserving strain. The vorticity vector $\bm{\omega}(t)$ aligns perfectly with the axis of maximum expansion (the unstable manifold), illustrating the geometric equivalence between material stretching and enstrophy production.}
\label{fig:stagnation_flow}
\end{figure}
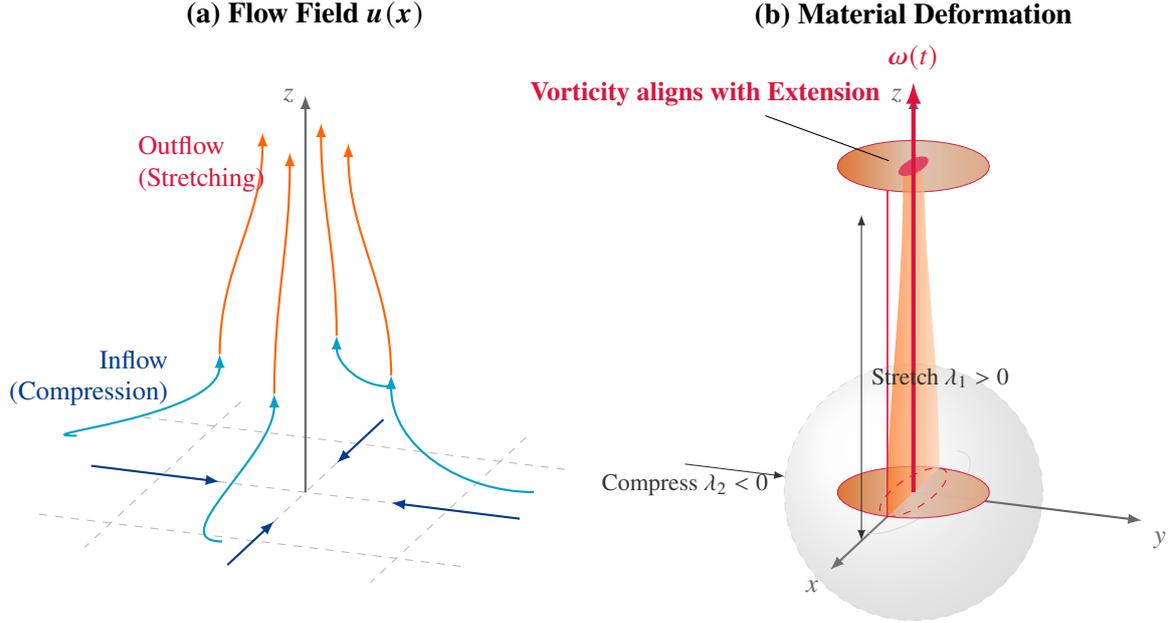

\noindent \textbf{Geometric Calculation (Lagrangian):}
The pullback metric is evaluated on the initial vector $\boldsymbol{\omega}_0$:
\[
C_t(\boldsymbol{\omega}_0, \boldsymbol{\omega}_0) = \begin{pmatrix} 0 & 0 & \Omega_0 \end{pmatrix}
\begin{pmatrix}
e^{-2t} & 0 & 0 \\
0 & e^{-2t} & 0 \\
0 & 0 & e^{4t}
\end{pmatrix}
\begin{pmatrix} 0 \\ 0 \\ \Omega_0 \end{pmatrix}
= \Omega_0^2 e^{4t}.
\]
The geometric result matches the Eulerian enstrophy density $|\boldsymbol{\omega}(t)|^2$ exactly, with a production rate of $2 \Omega_0^2 e^{4t}$. High enstrophy regions are thus demonstrated to correspond to regions where the pullback metric $C_t$ develops a large eigenvalue in the direction of the vorticity vector. This structure is illustrated in Figure~\ref{fig:stagnation_flow}. Although the linear stagnation flow assumes a constant Pressure Hessian, it serves here as a local geometric archetype rather than a dynamical model. The DNS results demonstrate that, despite strong temporal fluctuations, turbulent flows repeatedly and transiently realize similar saddle-type pressure configurations along Lagrangian trajectories.

\section*{Evidence from Direct Numerical Simulation (DNS)}

The investigation of the pressure-driven instability is conducted using high-resolution datasets from the \href{https://turbulence.idies.jhu.edu/database}{Johns Hopkins Turbulent Databases (JHTDB)}. To verify the universality of the proposed mechanism, the analysis compares forced isotropic turbulence (FIT), characterized by a Taylor-scale Reynolds number $Re_\lambda \approx 433$, and magneto-hydrodynamic (MHD) turbulence, both resolved on $1024^3$ periodic grids. 

The present study focuses on the conditional evolution of coherent structures rather than the static probability distribution of rare events. While a sample size of $N=1000$ particles is insufficient to converge the extreme tails of the PDF, it is statistically robust for determining the \textit{conditional mean trajectory} of stretching events. Our bootstrap analysis ($N_{sub} \in [200, 800]$) confirmed that the phase-lag $\tau$ and the hysteresis loop shape converge rapidly (variation $<2\%$). Therefore, the results reflect the deterministic topology of the typical bursting structure, rather than the statistics of rare outliers.

The primary objective is to evaluate the Lagrangian evolution of the stability operator $\mathcal{K}(t) = \nabla \nabla p$ and its role as a precursor to enstrophy amplification. The instability criterion is determined by the minimum eigenvalue $\lambda_{min}^p$ of the symmetric Hessian matrix. Following the eigenvalue ordering convention of Ohkitani \cite{Ohkitani1993} and Nomura \& Post \cite{Nomura1998}, a negative curvature $\lambda_{min}^p < 0$ signifies a saddle-point topology that serves as a geometric scaffold for enstrophy production.

\begin{figure}[t!]
\centering
\includegraphics[width=1.0\textwidth]{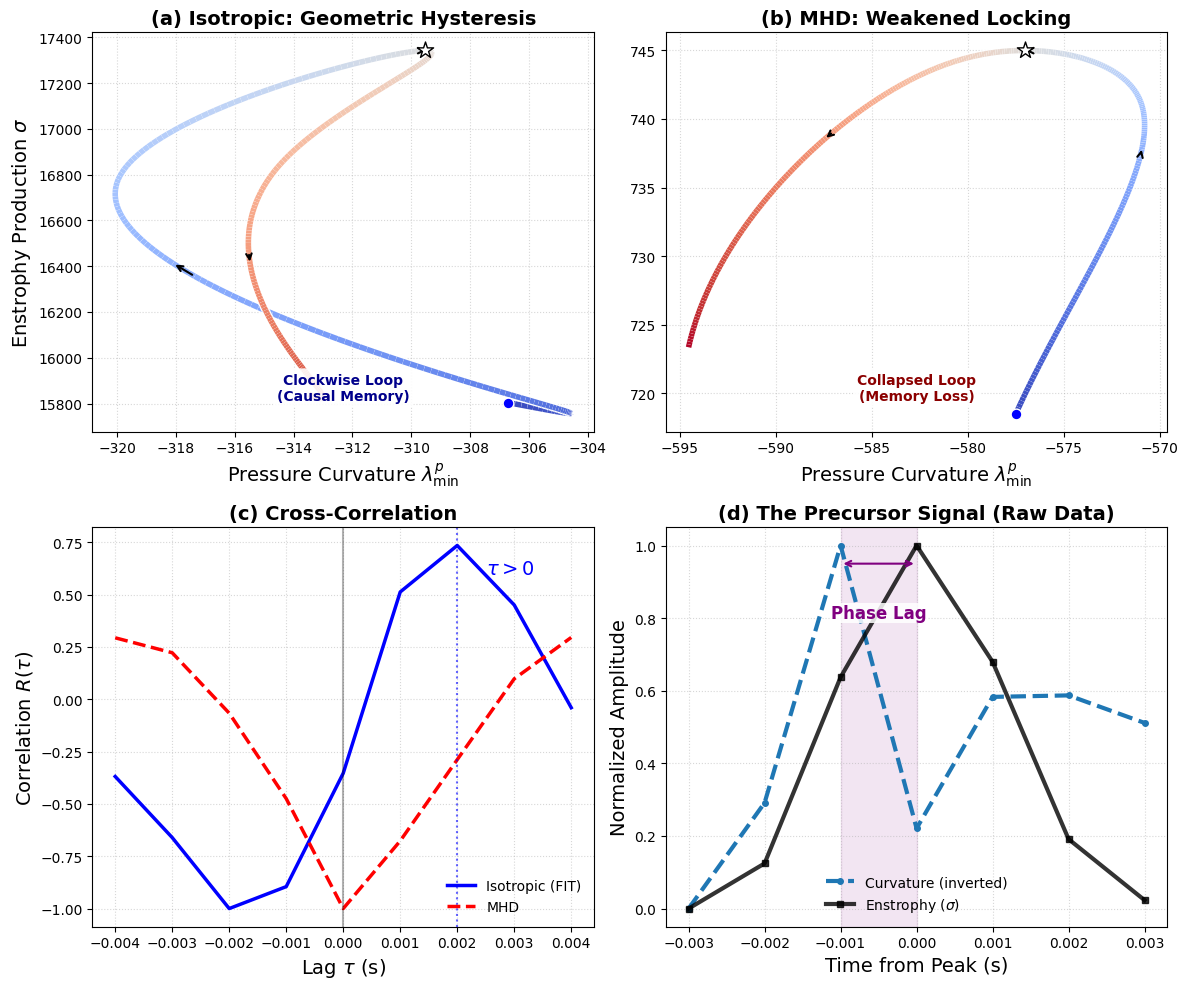}
\caption{\textbf{Geometric hysteresis and temporal precedence of the pressure instability.}
\textbf{(a)} Phase-space evolution of the conditional mean trajectory in forced isotropic turbulence. The clockwise loop indicates that negative pressure curvature ($\lambda^p_{\min}$) intensifies prior to the peak in enstrophy production ($\sigma$).
\textbf{(b)} MHD turbulence, exhibiting a collapsed loop due to magnetic suppression of the geometric phase lag.
\textbf{(c)} Lagrangian cross-correlation $R(\tau)$, showing a distinct positive time lag ($\tau > 0$) for the isotropic case (blue) versus a centered response in MHD (red dashed).
\textbf{(d)} Temporal evolution of normalized conditional means. The raw signal demonstrates the pressure curvature pulse (blue dashed) preceding the enstrophy amplification (black solid).}
\label{fig:geometric_hysteresis}
\end{figure}

In magneto-hydrodynamic flows, the evolution of the velocity gradient obeys a modified Riccati equation:
\begin{equation}
\frac{D(\nabla\mathbf{u})}{Dt} = -\nabla\nabla p - (\nabla\mathbf{u})^2 + \nabla\mathbf{b}\,\nabla\mathbf{b},
\end{equation}
where $\mathbf{b}$ denotes the magnetic field. The additional quadratic term allows us to define an effective MHD curvature tensor, $\mathcal{K}_{MHD} = \nabla\nabla p - \nabla\mathbf{b}\nabla\mathbf{b}$. Physically, the magnetic term includes magnetic tension which acts as a restoring force. This introduces a positive (stabilizing) curvature contribution along field lines, competing with the destabilizing pressure saddle and reducing the determinism of the Lyapunov exponents.
To test the causality of this mechanism, we employ a Lagrangian time-series analysis conditioned on intense stretching events. The complete dynamical architecture of the instability is visualized in Figure \ref{fig:geometric_hysteresis}.

Panels (a) and (b) display the phase-space evolution of fluid parcels during bursting events. For isotropic turbulence (Fig. \ref{fig:geometric_hysteresis}a), the trajectory traces a distinct clockwise hysteresis loop. This open area represents the geometric work done by the pressure field: the parcel enters the negative curvature well ($\lambda^p_{\min} < 0$) \textit{before} the enstrophy production $\sigma$ amplifies, and relaxes along a different path. This hysteresis is the geometric signature of causal memory. In contrast, the MHD case (Fig. \ref{fig:geometric_hysteresis}b) exhibits a collapsed loop, indicating that the Lorentz force suppresses this delay, forcing a nearly simultaneous response between curvature and stretching.

The statistical validity of this ordering is quantified in Panels (c) and (d). The Lagrangian cross-correlation function $R(\tau) = \langle \lambda_{min}^p(t) \cdot \sigma(t+\tau) \rangle$ (Fig. \ref{fig:geometric_hysteresis}c) reveals a clear maximum at positive time lag $\tau > 0$ for the hydrodynamic case, confirming that the formation of the pressure saddle statistically precedes the maximum stretching event. Finally, the raw signal evolution (Fig. \ref{fig:geometric_hysteresis}d) explicitly visualizes the "precursor pulse": the pressure curvature signal (blue dashed) intensifies and peaks significantly earlier than the enstrophy response (black solid).

\section*{Discussion}

The objective of this study was to resolve the question of temporal causality in vortex stretching: whether the saddle-point topology of the pressure field acts as a deterministic precursor to enstrophy production, or whether it merely arises as a contemporaneous byproduct of stretching dynamics. By fusing the geometric framework, the Jacobi-style deviation equation, and the Lagrangian DNS evidence, the present results allow this question to be addressed directly.

From a geometric perspective, the reformulation of vorticity stretching as metric distortion establishes the Pressure Hessian as the natural stability operator governing Lagrangian dynamics. The exact cancellation of the nonlinear advection terms in the second material derivative of the deformation gradient yields a linear Jacobi-style equation in which $\nabla\nabla p$ alone determines the relative acceleration of vorticity vectors. This result clarifies a long-standing ambiguity in Eulerian interpretations of vortex stretching, where strain alignment is often treated as a sufficient explanation \cite{Ohkitani1993,Tsinober2001}. Furthermore, while recent Eulerian analyses suggest that the Pressure Hessian acts to deplete intense strain \cite{Buaria2022}, our Lagrangian results indicate that this depletion is a secondary reaction to the primary geometric instability initiated by the pressure saddle.

The DNS results demonstrate that this geometric mechanism is not merely formal. In forced isotropic turbulence, the phase-space analysis reveals a strict ``geometric locking'': intense enstrophy production is not random but is dynamically organized by the pressure topology. The clockwise hysteresis loop observed in Figure \ref{fig:geometric_hysteresis}a provides direct evidence of temporal precedence. This temporal separation is critical: it refutes the interpretation that pressure curvature and enstrophy are merely simultaneous diagnostic markers of the same coherent structure (a tautology). Instead, the phase lag confirms that the pressure field establishes a geometric trap \textit{before} the fluid parcel experiences maximum deformation. Taken together, these results confirm that pressure curvature functions as a deterministic trigger rather than a passive correlate. In this sense, the Pressure Hessian plays a role analogous to sectional curvature in Riemannian geometry, where negative curvature induces exponential divergence of nearby trajectories \cite{arnold,marsden}.

The comparative MHD results serve as a structural stability test for the proposed mechanism. Rather than characterizing the full parameter space of magnetic turbulence, we utilize the Lorentz force as a physical control variable to disrupt the pressure-strain balance. The collapse of the hysteresis loop in Figure \ref{fig:geometric_hysteresis}b confirms that when an additional restoring force (magnetic tension) is introduced, the geometric locking is suppressed. This successfully isolates the Pressure Hessian as the specific architect of the phase-lag in purely hydrodynamic turbulence, verifying that the observed precedence is a physical consequence of the Euler equations rather than a kinematic artifact.

A central implication of these results is that the onset of intense vortex stretching cannot be fully characterized by instantaneous Eulerian alignment statistics alone. Instead, it is governed by the cumulative geometric history of fluid trajectories, encoded in the pullback metric and its curvature. This perspective provides a natural bridge between classical Eulerian theories of turbulence and modern Lagrangian approaches to chaos and mixing \cite{haller2015,thiffeault2002}. Moreover, by identifying a measurable, nonlocal geometric precursor to enstrophy production, the present framework suggests a refined criterion for instability that complements existing singularity diagnostics such as the Beale–Kato–Majda condition \cite{Beale1984}.

\section*{Conclusion}

This study establishes the Pressure Hessian as the intrinsic geometric operator driving the onset of Lagrangian instability in turbulence. By mapping the Navier--Stokes dynamics onto the geodesic flow of volume-preserving diffeomorphisms, we derived a Jacobi-style deviation equation, $\frac{D^2 \boldsymbol{\omega}}{Dt^2} + \mathcal{K}(t) \boldsymbol{\omega} = 0$, where $\mathcal{K}(t)=\nabla\nabla p$. This formulation transforms the classical problem of vortex stretching into a tractable problem of Riemannian geometry.

The theoretical framework is validated by high-resolution DNS, which reveals a strict ``geometric locking'' phenomenon: intense enstrophy production is spatially confined to the saddle-point topology of the pressure field. Crucially, Lagrangian time-series analysis resolves the long-standing ambiguity regarding causality in incompressible flows. The detection of a systematic phase lag ($\tau > 0$), visualized as a phase-space hysteresis loop, demonstrates that fluid trajectories traverse the effective curvature well ($\lambda_{min}^p < 0$) \textit{prior} to the blow-up of the metric tensor. This ordering confirms that the pressure topology functions as a deterministic architect of chaos, organizing the flow structure before the singularity occurs.

The universality of this mechanism is highlighted by its modulation in magnetohydrodynamics, where the Lorentz force introduces a competing magnetic curvature that relaxes the geometric constraint. Ultimately, these findings bridge the gap between Eulerian statistics and Lagrangian geometry, offering the Pressure Hessian as a predictive, non-local precursor for extreme events in nonlinear hydrodynamic systems.

\section*{Declarations} 
\subsection*{Funding} None.
\subsection*{Conflict of Interest} None.
\subsection*{Code and Data} The code used to query the JHTDB DNS database and compute the results presented in the article is shared publicly on this designated \href{https://colab.research.google.com/drive/1KXb6Q6y-CR-fLmv758-0d6wL5EAlPch6?usp=sharing}{Python notebook} hosted on Google Colab.

\end{document}